\documentclass[12pt,preprint,showpacs,preprintnumbers]{revtex4}
\usepackage{amsfonts}
\usepackage{amsmath}
\usepackage{amssymb}
\usepackage{hyperref}
\usepackage{color}
\usepackage{graphicx}
\usepackage{amssymb}
\usepackage{amsmath}
\usepackage{graphicx}
\usepackage{dcolumn}
\usepackage{bm}
\usepackage{epsfig}
\usepackage[T1]{fontenc}
\usepackage{ae,aecompl}

\begin{document}

\title{Simulation study on cavity growth in ductile metal materials under dynamic loading}
\author{Aiguo Xu\footnote{Corresponding author. E-mail address: Xu\_Aiguo@iapcm.ac.cn},
Guangcai Zhang, Yangjun Ying, and Xijun Yu}
\address{National Key Laboratory of Computational Physics, \\
Institute of Applied Physics and Computational Mathematics, P. O.
Box 8009-26, Beijing 100088, P.R.China}

\date{\today}

\begin{abstract}
Cavity growth in ductile metal materials under dynamic loading is investigated via the material point method. Two typical cavity effects in the region subjected to rarefaction wave are identified:
(i) part of material particles flow away from the cavity in comparison to the initial loading velocity,
(ii) local regions show weaker negative or even positive pressures.
Neighboring cavities interact via coalescence of isobaric contours. The growth of cavity under tension shows staged behaviors. After the initial slow stage, the volume and the dimensions in both the tensile and transverse directions show linear growth rate with time until the global tensile wave arrives at the upper free surface. It is interesting that the growth rate in the transverse direction is faster than that in the tensile direction. The volume growth rate linearly increases with the initial tensile velocity. After the global tensile wave passed the cavity, both the maximum particle velocity in the tensile direction and the maximum particle velocity in the opposite direction  increase logarithmically with the initial tensile speed. The shock wave reflected back from the cavity and compression wave from the free surface induce the initial behavior of interfacial instabilities such as the Richtmyer-Meshkov instability, which is mainly responsible for the irregularity in the morphology of deformed cavity. The local temperatures and distribution of hot spots are determined by the plastic work. Compared with the dynamical process, the heat conduction is much slower.

\end{abstract}


\vspace{2pc}
\maketitle

%%%%%%%%%%%%%%%%%%%%%%%%%%%%%%%%%%%%%%%%%%%%%%%%%%%%%%%%%%%%%%%

\section{Introduction}

Failure of ductile metal materials under dynamic loading is an important and fundamental issue in the fields of science and technology. The failure process is complicated because it couples various physical and mechanical mechanisms in the microscopic, mesoscopic and macroscopic scales. Globally speaking, spallation or fragmentation of metal material is mainly  composed of the following typical stages, nucleation, growth and coalescence of microscopic voids and/or larger scale cavities. There have been extensive studies on the quasistatic growths of voids and cavities. The dynamical growth behaviors are much more complicated and far from being well understood\cite{1}.

In 1972 Carroll and Holtz\cite{42} studied the static and dynamic cavity-collapse relations for ductile porous materials and found that the compression effect on the cavity growth is not pronounced when the material is not sensitive to the loading rate. The research was extended to the visco-plastic materials by Johnson\cite{43} in 1981.
In 1987 Becker\cite{Becker1987} numerically analyzed the effect of a nonuniform distribution of porosity on flow localization and failure in a porous material. The void density distribution and properties used to characterize the material behavior were obtained from measurements on partially consolidated and sintered iron powder. The calculations were carried out using an elastic viscoplastic constitutive relation for porous plastic solids. Local material failure is incorporated into the model through the dependence of the flow potential on void volume fraction. The region modeled is a small portion of a larger body under various stress conditions. Both plane strain and axisymmetric deformations are considered with imposed periodic boundary conditions. It was found that interactions between regions with higher void fractions promote plastic flow localization into a band, and that local failure occurs via void growth and coalescence within the band. The results of this study suggested a failure criterion based on a critical void volume fraction that is only weakly dependent on stress history. The critical void fraction depends on the initial void distribution and material hardening characteristics.
In 1992 Ortiz and Molinari\cite{44} studied the effect of strain hardening and rate sensitivity on the dynamic growth of a void
in a plastic material and pointed out that the inertial effect, hardening effect, loading rate effect can significantly influence the void growth.  The studies of Benson\cite{45} in 1993 and of Ramesh and Wright\cite{46} in 2003 showed that the inertia effect is responsible for stable growth of the cavity.
In 1998 Pardoen, et al.\cite{Pardoen1998} investigated the ductile fracture of round copper bars within the scope of the local approach methodology. Two damage models,  the Rice-Tracey model and the Gurson-Leblond-Perrin model, were analyzed. Four coalescence criteria, (i) a critical value of the damage parameter, (ii) the Brown and Embury criterion, (iii) the Thomason criterion and (iv) a criterion based on the reaching of the maximum von Mises equivalent stress in a Gurson type simulation, were comparatively studied. Ellipsoidal void growth and void interaction were accounted for. As far as possible, all the parameters of the models were identified from experiments and physical observations. The effect of stress triaxiality was studied using specimens presenting a wide range of notch radii. The effect of strain-hardening was analyzed by comparing the behaviour of the material in the cold drawn state and in the annealed state.
In 2000 Pardoen et al.\cite{Pardoen2000} proposed an extended model for void growth and coalescence. This model integrated two existing contributions, the Gologanu-Leblond-Devaux model extending the Gurson model to void shape effects and the Thomason scheme for the onset of void coalescence. Each of these was extended heuristically to account for strain hardening. In addition, a micromechanically-based simple constitutive model for the void coalescence stage was proposed to supplement the criterion for the onset of coalescence. The fully enhanced Gurson model depends on the flow properties of the material and the dimensional ratios of the void-cell representative volume element. It incorporates the effect of void shape, relative void spacing, strain hardening, and porosity.
In 2001 Orsini, et al.\cite{Orsini2001}
developed an inelastic rate-dependent crystalline constitutive formulation and specialized computational schemes  to obtain a detailed understanding of the interrelated physical mechanisms which can result in ductile material failure in rate-dependent porous crystalline materials subjected to finite inelastic deformations.
Results of this study are consistent with experimental observations that ductile failure can occur either due to void growth parallel to the stress axis, which results in void coalescence normal to the stress axis, or void interaction along bands, which are characterized by intense shear-strain localization and that intersect the free surface at regions of extensive specimen necking.
In 2002 Tvergaard and and Hutchinson\cite{2002a} discussed two mechanisms of ductile fracture, void by void growth versus multiple void interaction, Zohdi, et al.\cite{2002b} discussed
the plastic flow in porous material.

Currently, most of the studies on cavity/void growth are focused on their relevance on macroscopic behaviors\cite{42,43,Becker1987,44,45,46,Pardoen1998,Pardoen2000,Orsini2001,2002a,2002b}. The quantitative relations are obtained by fitting experimental results. Those studies do not reveal or indicate the underlying idiographic physical and mechanical mechanisms of cavity/void growth. Cavity coalescence is the final stage of spallation developed from mesoscopic scale to macroscale\cite{27}. It is also the least-known stage\cite{47,48,Zurek1998,49,50,51,52,53}. Continuous damage mechanics adopts fluid or solid description supplemented by damage modeling. The damage is generally modeled by an internal variable. The internal variable is defined by the variation of some mechanical behavior and is not dynamically relevant to the particular structures.

The molecular dynamics simulations\cite{47,PRB2005,MSMSE2006,Pang1,Pang2} can help understand some mechanisms from the atomic scale, but the temporal and spatial scales it can access are too small to be comparable with experiments.
The Material Point(MP) method\cite{MPM1,MPM2,MPM4,MPM5} is a newly developed mesoscopic particle method in the field of computational solid mechanics. In this method, the continuum bodies are discretized with $N$ material particles. Each material particle carries the information of position, velocity, temperature, mass, density, Cauchy stress, strain tensor and all other internal state variables necessary for the constitutive model.  At each time step, calculations consist of two parts: a Lagrangian part and an Eulerian  one. Firstly, the material particles flow
with the body, and is used to determine the strain increment, and the stresses in the sequel. Then, the velocity field is mapped from the particles to the Eulerian mesh nodes. The spatial derivatives are calculated and the momentum equation is integrated. The velocity and acceleration fields are mapped back to update those of the particles\cite{CTP2008,JPD2008}. The MP method not only takes advantages of both the Lagrangian and Eulerian algorithms but avoids their drawbacks as well. Since using Eulerian background grid,  it is more stable and has a higher computational efficiency than the meshless smooth particle hydrodynamics(SPH)\cite{MPM5}.

 From the physics side, the MP method is based on continuum medium description and designing of contact force. It has been extensively used to  simulate the complex dynamical behaviors of shock wave interaction on inhomogeneous materials\cite{JPCM2007,CTP2009,CTP2009b,SciCN2010,PhysScr2010,CAMWA2011}. The mesoscopic MP simulation can be further used to investigate the growth and evolution of defect structures such as cavities and cracks in the scales of micron and larger. Such investigations may present indicative results for improving physical modeling of fracture in larger scales. From the simulation side, in the MP method the continuous portion and the cavities are considered separately. So, it is convenient to set the particular structures according to our need and convenient to obtain the concrete information on the shapes, sizes, connectivity of relevant structures and their influences on surrounding materials. In other words, it is convenient to recover with more fidelity the physical processes of damage and failure. Simulation results may work as theoretical bases for the physical modeling of damnification. Different from the phenomenological quasistatic analysis, the MP simulation results contain intrinsically the inertial effects.

\section{Theoretical model of the material}

We assume that the material follows an associative von Mises
plasticity model with linear kinematic and isotropic hardening\cite{CModel}.
The stress and strain tensors, $\boldsymbol{\sigma }$
and $\boldsymbol{\varepsilon }$, reads
$
\boldsymbol{\sigma } = \mathbf{s}-P\mathbf{I}
$, $
P=-\verb|Tr|(\boldsymbol{\sigma })/3
$, $
\boldsymbol{\varepsilon } = \mathbf{e}+\theta \mathbf{I}/3
$, $
\theta =\verb|Tr|(\boldsymbol{\varepsilon })/3
$,
where $P$ is the pressure scalar, $\mathbf{s}$ the deviatoric stress tensor,
and $\mathbf{e}$ the deviatoric strain. The strain $\mathbf{e}$ is generally
decomposed as $\mathbf{e}=\mathbf{e}^{e}+\mathbf{e}^{p}$, where $\mathbf{e}%
^{e}$ and $\mathbf{e}^{p}$ are the traceless elastic and plastic components,
respectively. The material shows a linear elastic response until the von
Mises yield criterion is reached. The yield $\sigma _{Y}$ increases linearly with the second invariant of the plastic
strain tensor $\mathbf{e}^{p}$, i.e.,
$
\sigma _{Y}=\sigma _{Y0}+E_{Harden}\left\Vert \mathbf{e}^{p}\right\Vert
$,
where $\sigma _{Y0}$ is the initial yield stress and $E_{Harden}$ the
hardening coefficient. The deviatoric stress $\mathbf{s}$ is related to the
Young's module $Y$, the Poisson's ratio $\nu $ and the traceless elastic stress tensor by
$
\mathbf{s}=\mathbf{e}^{e} Y/(1+\nu)
$.
The shock speed $U_{s}$ and the particle speed $U_{p}$ after
the shock follows a linear relation, $U_{s}=c_{0}+\lambda U_{p}$, where $c_{0}$ is the sound speed and
$\lambda $ a characteristic coefficient of material. The pressure $P$ is
calculated by the following Mie-Gr\"{u}neisen state of equation\cite{explosion},
\begin{equation}
P-P_{H}=\frac{\gamma (V)}{V}[E-E_{H}(V_{H})]\mathtt{,}  \label{eq-eos}
\end{equation}%
where
$P_{H}$, $V_{H}$ and $E_{H}$ are pressure, specific volume and energy on the
Rankine-Hugoniot curve, respectively. The relation between $P_{H}$ and
$V_{H} $ can be written as
\begin{equation}
P_{H}=\left\{
\begin{array}{ll}
\frac{\rho _{0}c_{0}^{2}(1-\frac{V_{H}}{V_{0}})}{(\lambda -1)^{2}(\frac{%
\lambda }{\lambda -1}\times \frac{V_{H}}{V_{0}}-1)^{2}}, & V_{H}\leq V_{0}
\\
\rho _{0}c_{0}^{2}(\frac{V_{H}}{V_{0}}-1), & V_{H}>V_{0}%
\end{array}%
\right.
\end{equation}
The increment  of specific internal energy $E-E_{H}(V_{H}) $ is taken as the plastic energy. Both the shock compression
and the plastic work contribute to the increasing of temperature. The
temperature increment from shock compression is calculated by
\begin{equation}
\frac{\mathrm{d}T_{H}}{\mathrm{d}V_{H}}=\frac{c_{0}^{2}\cdot \lambda
(V_{0}-V_{H})^{2}}{c_{v}\big[(\lambda -1)V_{0}-\lambda V_{H}\big]^{3}}-\frac{%
\gamma (V)}{V_{H}}T_{H}.  \label{eq-eos-temprshock}
\end{equation}%
where $c_{v}$ is the specific heat. The increasing of temperature from plastic work is
$
\mathrm{d}T_{p}=\mathrm{d}W_{p}/c_{v}
$.

In this work we choose aluminum as the sample material. The
corresponding parameters are as below: initial material density in the solid portion $\rho_{0}=2700$ kg/m$^{3}$, $Y=69$ Gpa,
$\nu =0.33$, $\sigma _{Y0}=120$ Mpa, $E_{Harden}=384$ MPa,
$c_{0}=5.35$ km/s, $\lambda=1.34$, $c_{v}=880$ J/(Kg$\cdot $K),
$k=237$ W/(m$\cdot $K) and $\gamma_0=1.96$ when the pressure is
below $270$ GPa. The initial temperature of the material is 300 K.

%%%%%%%%%%%%%%%%%%%%%%%%%%%%%%%%%%%%%%%%%%%%%%%%%%%%%%%%%%%%%%%
%%%%%%%%%%%%%%%%%%%%%%%%%%%%%%%%%%%%%%%%
\begin{figure}[tbp]
\center\includegraphics*%
[bbllx=0pt,bblly=0pt,bburx=533pt,bbury=476pt,angle=0,width=0.85\textwidth]
{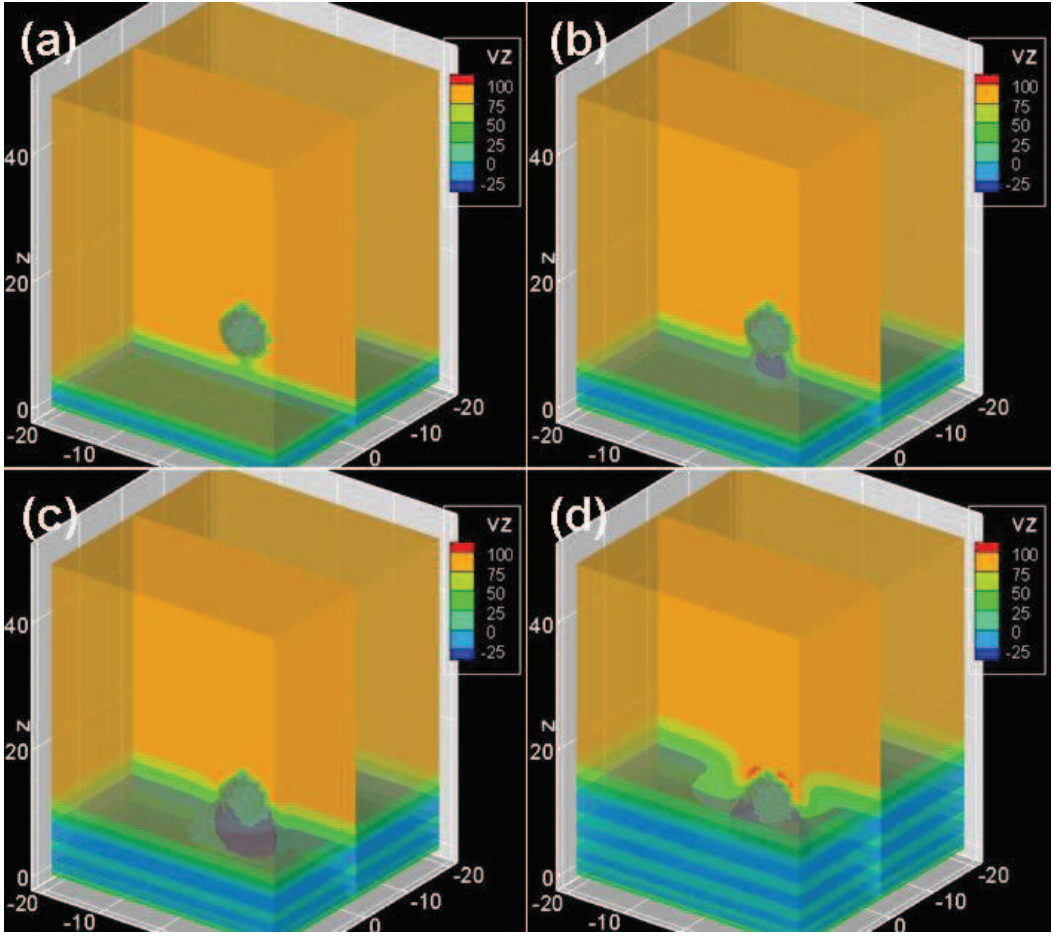}
\caption{Configurations with $v_z$ field at four different times for the case with $z=10 \mu$m and initial $v_{z0}=100$m/s. The contours for $v_z=0$ are shown in the plots. (a)$t=0.8$ns, (b)$t=1.2$ns, (c)$t=2.0$ns and (d)$t=3.0$ns. }
\end{figure}
%%%%%%%%%%%%%%%%%%%%%%%%%%%%%%%%%%%%%%%%%

%%%%%%%%%%%%%%%%%%%%%%%%%%%%%%%%%%%%%%%%
\begin{figure}[tbp]
\center\includegraphics*%
[bbllx=0pt,bblly=0pt,bburx=583pt,bbury=497pt,angle=0,width=0.85\textwidth]
{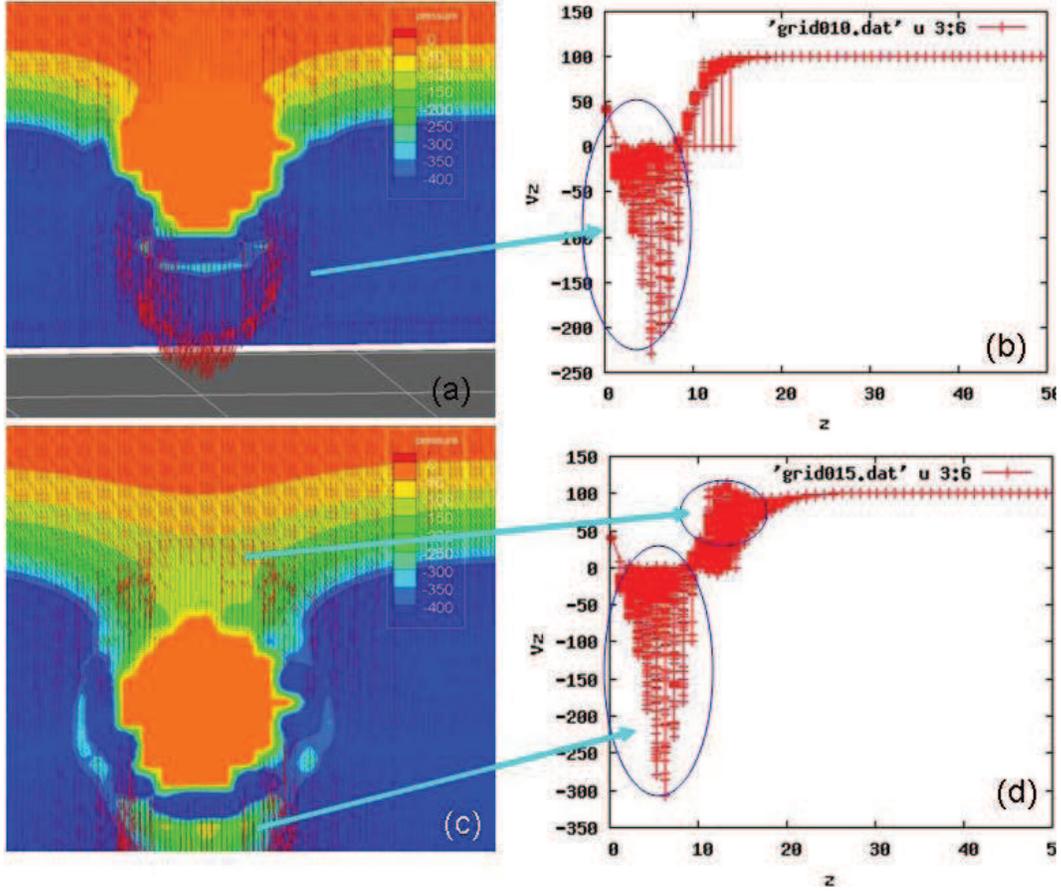}
\caption{Configurations with pressure and velocity fields in the plane with $x=0$ [see (a) and (c)] and $v_z$ distribution in the tensile direction [see (b) and (d)]. $t=2$ns in (a) and (c). $t=3$ns in (b) and (d).}
\end{figure}
%%%%%%%%%%%%%%%%%%%%%%%%%%%%%%%%%%%%%%%%%
%%%%%%%%%%%%%%%%%%%%%%%%%%%%%%%%%%%%%%%%
\begin{figure}[tbp]
\center\includegraphics*%
[bbllx=0pt,bblly=0pt,bburx=540pt,bbury=320pt,angle=0,width=0.85\textwidth]
{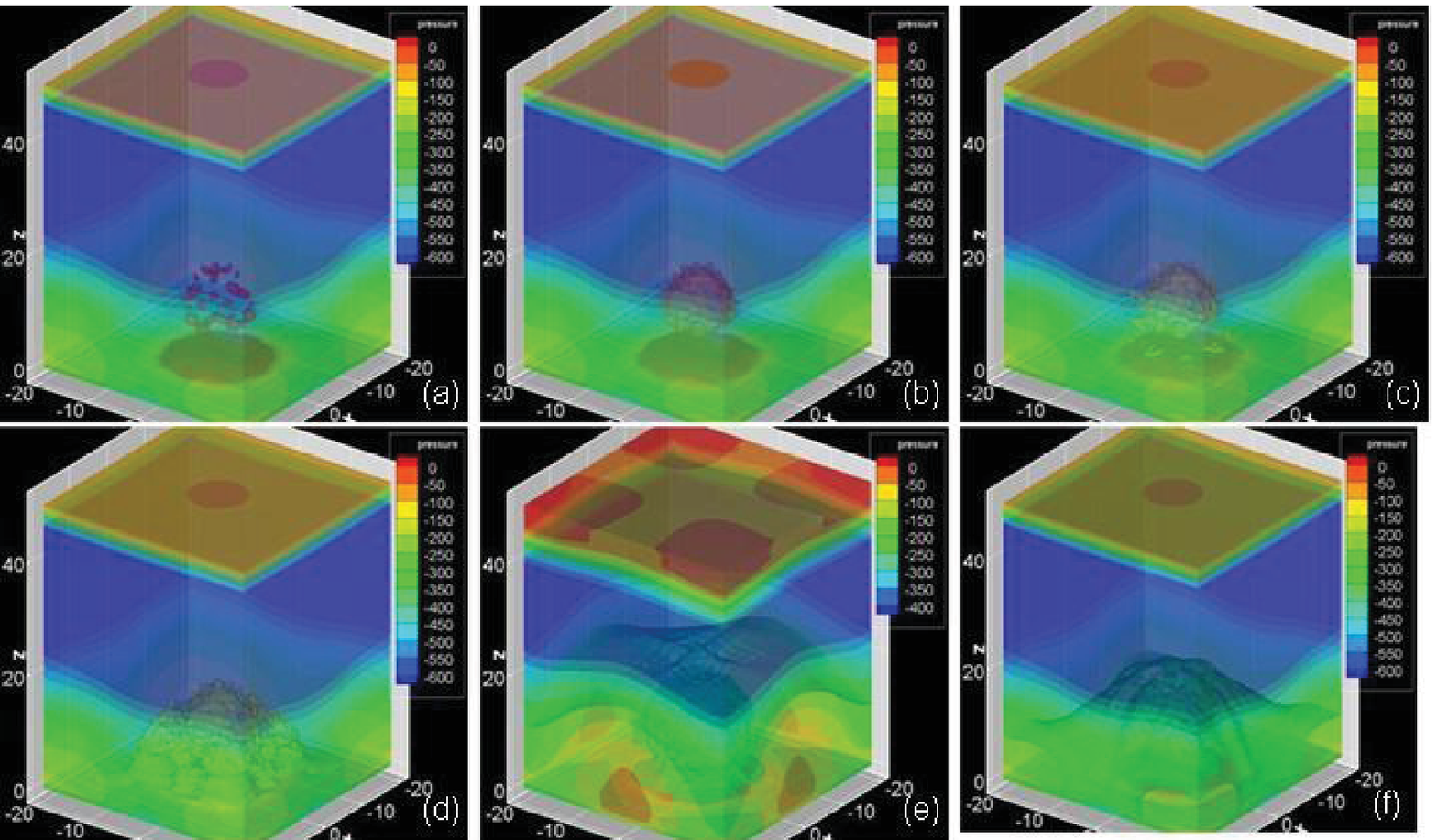}
\caption{Configurations with pressure field at the time $9$ns. The pressure contours in (a)-(f) correspond to $-300$MPa, $-350$MPa, $-400$MPs, $-450$MPa, $-500$MPa, and $-550$MPa, respectively.}
\end{figure}
%%%%%%%%%%%%%%%%%%%%%%%%%%%%%%%%%%%%%%%%%

%%%%%%%%%%%%%%%%%%%%%%%%%%%%%%%%%%%%%%%%%%%%%%%%%%%%%%%%%%%%%%%
\section{Simulation results and physical interpretation}

\subsection{Global scenario}
In our simulations the body of aluminum material with cavity is connected with a rigid wall fixed at the bottom with the coordinate $z=0$. The simulated body is located within the volume, $[-20,20] \times [-20,20] \times [0,50]$ with the length unit $\mu$m. Initially, a spherical cavity with radius $r=5$ $\mu$m is located at the position $(0,0,z)$ within the material body. At the time $t=0$ the material body with cavity starts to move upward at the velocity $v_{z0}$. Thus, the rarefaction wave or tensile wave occurs at the plane with $z=0$. The rarefaction wave propagates upwards within the material body. In our MP simulations, the mesh size is $1 \mu$m and the diameter of the material particle is $0.5 \mu$m. Periodic boundary conditions are used in the horizontal directions and free boundary condition is used in the upper surface of the material body. The rigid wall is assumed to be the same kind of material with the material body. Therefore, no special treatment of contact surfaces is needed in the MP simulations.

Figure 1 shows the snapshots of configurations with $v_z$ field at four different times for the case with $z=10 \mu$m and initial $v_{z0}=100$m/s. Figures.1(a)-1(d) correspond to the times $t=0.8$ns, $1.2$ns, $2.0$ns and $3.0$ns, respectively. The contours for $v_z=0$ are shown in the plots. Since no material particles are located within the cavity, the velocities at the nodes within the cavity are equal zero. Before the arrival of the global rarefaction wave, the upper contour with $v_z=0$ presents the initial morphology of the cavity. In Figs.1(a)-1(d) the moving upwards of the lower $v_z=0$ contour shows the propagation of rarefaction wave. It is clear from Fig.1(a) that the lower $v_z=0$ contour is approaching the lower boundary of the cavity at the time $t=0.8$ns. The velocities of particles below the cavity had begun to decrease before $t=0.8$ns. Below the lower $v_z=0$ contour some material particles show negative velocities. With propagating of the rarefaction wave, the lower $v_z=0$ contour begins to get connection with that corresponding to the cavity. When the rarefaction wave arrives at the cavity, compression wave is reflected back. Under the action of the reflected compression wave, more material particles show negative velocities and their amplitudes continue to increase.[See Figs.1(b) and 1(c).] The deformation rate of the cavity is slower compared with the propagation speed of rarefaction wave surrounding the cavity. After passing the cavity the surrounding rarefaction waves begin to converge. Therefore, stronger negative pressure appears on the top of the cavity. Material particles on the top of the cavity are accelerated by the upward stresses. At the time $t=3$ns some material particles show velocities larger than $100$m/s. See Fig.1(d).

Figures 2(a) and 2(c) show the configurations with pressure and velocity fields within the plane $x=0$ at times $t=2$ns and $t=3$ns, respectively. To investigate the amplitudes of particle velocities we show the distribution of $v_z$ along the vertical direction in the other plots of Fig.2. Figures 2(b) and 2(d) correspond to Figs.2(a) and 2(c), respectively. At the times $t=2$ns and $t=3$ns, the maximum downward or minimum particle velocities are $-230$m/s and $-300$m/s, respectively. From Fig.2(d) we can observe the vertical distribution of material particles with velocities larger than $100$m/s. In Figs.2(a) and 2(c) the color from blue to red corresponds to the increase of pressure. From Figs.2(a) and 2(c), one can observe the deformation of the cavity under tensile loading. The irregularities in the morphology of the cavity result from the following three aspects. (i) The initial cavity represented by the placed particles is not strictly spherical. (ii) The shock waves reflected back from the cavity induce the well-known Richtmyer-Meshkov(RM) hydrodynamic interfacial instability. The RM instability is the main mechanism for the initial irregularities of the deformed cavity. (iii) Compared with the dimension of the cavity, the mesh size is not small. For point (iii), it should be commented that if decrease the mesh unit, the body size can be simulated becomes smaller. We have to make compromise between the simulated body size and the mesh unit. It should also be pointed out that the practical cavities in materials are generally not strict spherical, which is qualitatively accordance with the simulated one.

With increasing of upward stress on the top of the cavity, the accelerations and velocities of particles within this region become larger. At the time $t=7.2$ns, the global rarefactive wave arrives at the upper free surface. The maximum velocity of particles on the top of the cavity is about $430$m/s. At this moment, there exists a region where the particles have large downward velocities below the cavity. The largest downward velocity is about $-325$m/s. In the plot of $v_z$ versus $z$, there is a valley between the peak and the rarefaction wave front. The smallest particle velocity is about $6$m/s. When the rarefaction wave arrives at the upper free surface and compression wave is reflected back. Within the region scanned by the reflected compression wave, material particles obtain downward accelerations. Several characteristics are typical for the unloading of rarefaction wave and reflecting back of compression wave. The first is the decreasing of velocities of material particles representing the upper free surface. The second is that the valley continues to move toward the upper free surface. The third is that the maximum velocity between the valley and the cavity continues to increase. At the same time, the region with maximum downward particle velocity moves toward the bottom. Since we use periodic boundary conditions in the horizontal directions, the simulation results for the case with  single cavity are also indicative for interaction of neighboring cavities. From the pressure field, it is clear that the negative pressures within regions among the neighboring cavities are weaker. The contours of negative pressure with small amplitudes get connection. The strength of compression wave reflected back from the cavity increases with increasing the strength of rarefaction wave. Consequently, local positive pressures occur among the neighboring cavities. The occurrence of positive pressures within the region scanned by the global rarefaction wave is a typical cavity effect.

Before the reflected compression wave arrive at the cavity, the deformation of the cavity is still controlled by the tensile loading. Below, we discuss the pressure distributions within the material at two times, 9 ns and 11 ns. Figure 3 show the configurations with pressure field at the time 9ns. The pressure contours in Figs. (a)-(f) are for $-300$Mpa, $-350$Mpa, $-400$Mpa, $-450$Mpa, $-500$Mpa and $-550$Mpa, respectively. Figure 3 shows that the contours around the cavities for pressure lower than -300Mpa are connected. The neighboring cavities get interaction via the connection of pressure contours. At the time 9ns, there is still no positive pressure occur among the neighboring cavities. Figure 4 shows various pressure contours at the time 11 ns. The contours in Figs. (a)-(f) are for $0$Mpa, $-50$Mpa, $-100$Mpa, $-150$Mpa, $-200$MPa and $-250$Mpa, respectively. Pressure distribution around the cavity is as below. (i) The pressure surrounding the cavity is zero. (ii) With increasing the pressure, the corresponding contour moves away from the cavity and its surface area becomes larger. (iii) Among cases shown in the figure the contour for $-150$Mpa has the maximum area. If further increase the pressure, the contour area becomes smaller. Pressure distribution between the cavity and rigid wall is as below. There are four regions around the cavity show positive pressure. The pressure contours for -100Mpa between the nearest cavities are connected. The contours for $-150$Mpa have a higher connectivity. All contours for $-200$Mpa, $-250$Mpa, etc. are connected. The pressure distribution on the top of the cavity is as below. The region with the highest pressure does not locate above the cavity but above the middle of neighboring cavities.  Since the rarefaction wave propagates more quickly within the solid region, the wave firstly arrives at the upper free surface and get reflected. The weaker the negative pressure, the closer to the upper free surface the corresponding contour, and the planar the corresponding contour.

%%%%%%%%%%%%%%%%%%%%%%%%%%%%%%%%%%%%%%%%
\begin{figure}[tbp]
\center\includegraphics*%
[bbllx=0pt,bblly=0pt,bburx=547pt,bbury=316pt,angle=0,width=0.85\textwidth]
{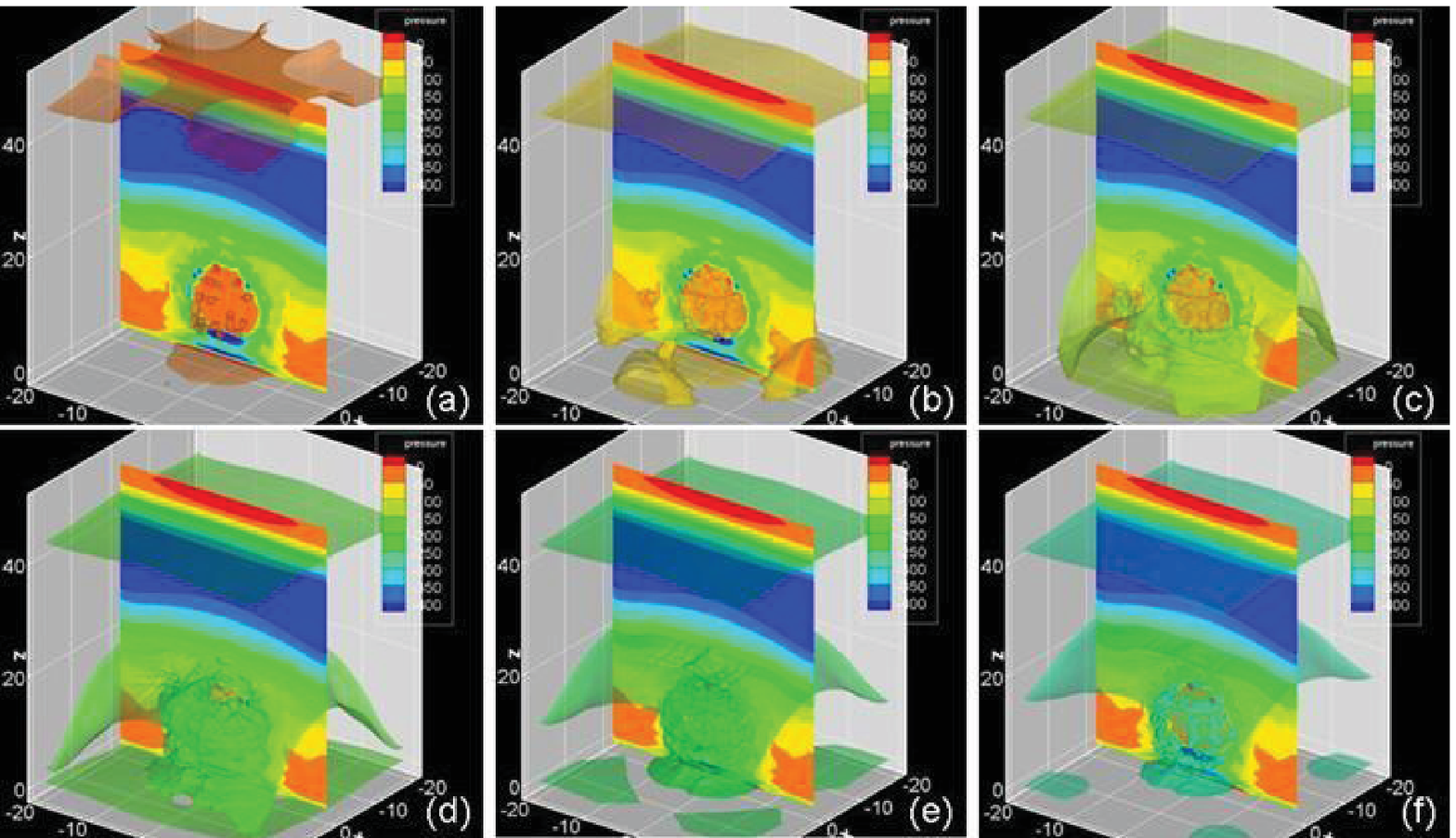}
\caption{Configurations with pressure field at the time $11$ns. The pressure contours in (a)-(f) correspond to $0$MPa, $-50$MPa, $-100$MPs, $-150$MPa, $-200$MPa, and $-250$MPa, respectively.}
\end{figure}
%%%%%%%%%%%%%%%%%%%%%%%%%%%%%%%%%%%%%%%%%
%%%%%%%%%%%%%%%%%%%%%%%%%%%%%%%%%%%%%%%%
\begin{figure}[tbp]
\center\includegraphics*%
[bbllx=0pt,bblly=0pt,bburx=536pt,bbury=476pt,angle=0,width=0.85\textwidth]
{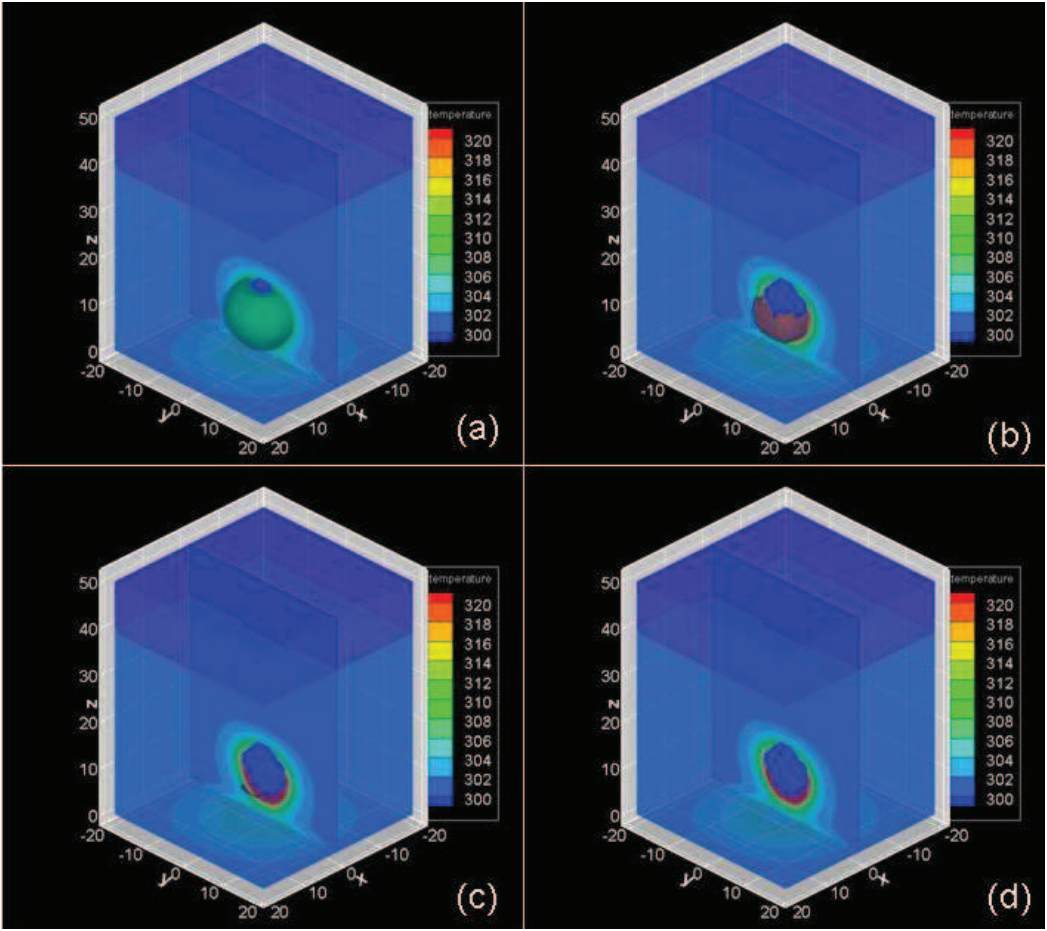}
\caption{Configurations with temperature field at the time $6$ns. The contours in (a)-(d) correspond to 310K, 320K, 330K and 340K, respectively.}
\end{figure}
%%%%%%%%%%%%%%%%%%%%%%%%%%%%%%%%%%%%%%%%%

%%%%%%%%%%%%%%%%%%%%%%%%%%%%%%%%%%%%%%%%
\begin{figure}[tbp]
\center\includegraphics*%
[bbllx=0pt,bblly=0pt,bburx=550pt,bbury=749pt,angle=0,width=0.70\textwidth]
{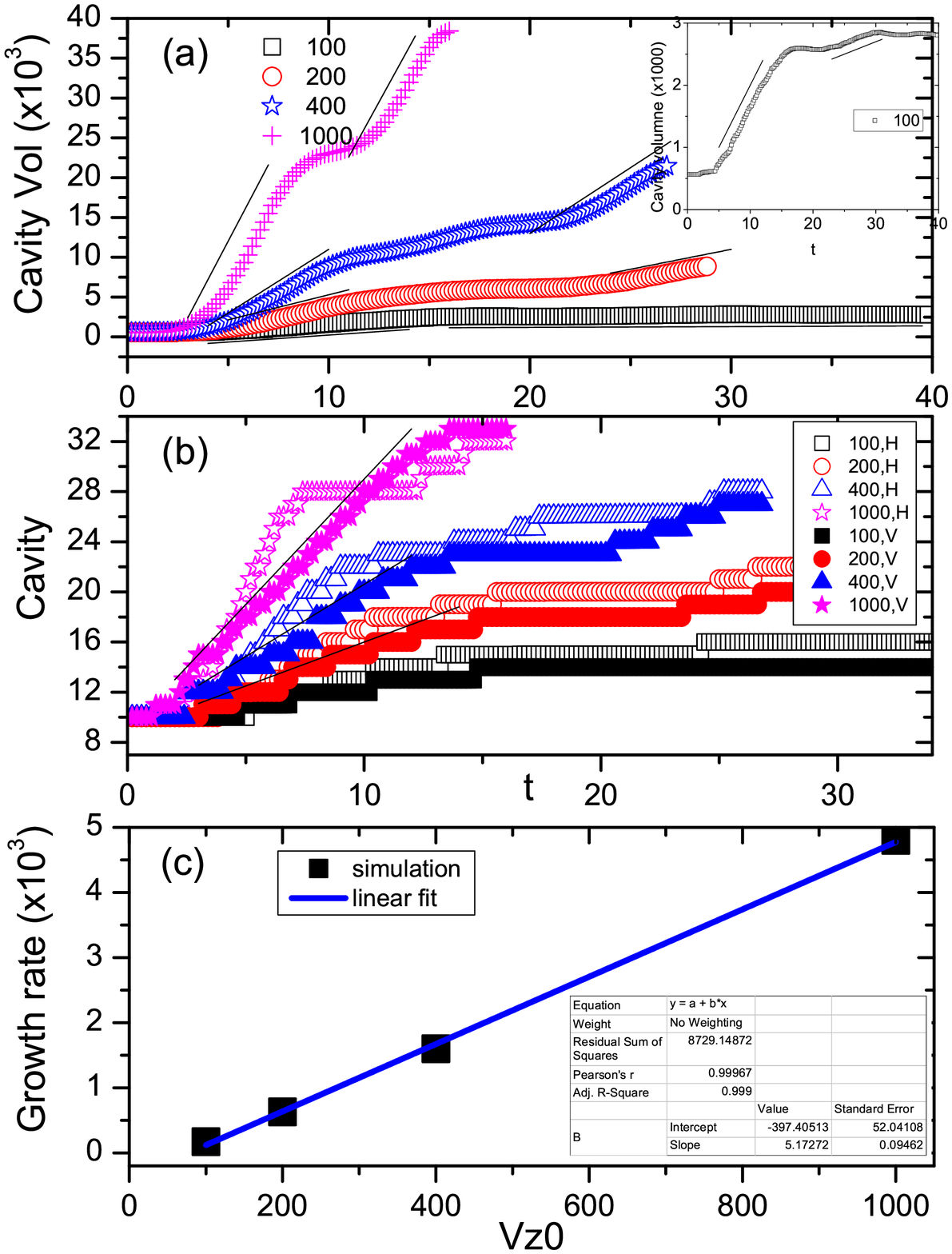}
\caption{Evolution of the cavity morphology. (a) Cavity volume versus time. (b) Cavity dimensions in the Horizontal(H) and Vertical(V) directions versus time. (c) The linear growth rate versus initial tensile velocity. The sizes of the initial tensile velocity $v_{z0}$, 100, 200, 400 and 1000, are shown in the legend of Fig.(a). The unit is m/s.  In Figs.(a) and (b) the points are simulation results and the lines are plotted to guide the eyes. An enlarge portion of the curve for $v_{z0}=100$ is shown in the inset of Fig.(a). In Fig.(c) the points are for the slopes of fitting lines in Fig. (a) for the first linear growth stage, and the line are linear fitting result for the points. }
\end{figure}
%%%%%%%%%%%%%%%%%%%%%%%%%%%%%%%%%%%%%%%%%
\subsection{Morphology versus tensile strength}
Figure 5 shows the configurations with temperature field at the time $6$ns. The contours in (a)-(d) are for $310$K, $320$K, $330$K and $340$K, respectively. Compared with the dynamical process, the thermal process is much slower. The temperature and distribution of hot-spots are mainly determined by the corresponding plastic work.

Since the rarefaction wave propagates in the sound speed, all rarefaction waves reach the upper free surface at the same time. With increasing the tensile strength, the growth rate of the cavity increases. Figure 6 shows the evolution of the cavity morphology. Figure 6(a) shows the cavity volume versus time.
The points are simulation results and the lines are plotted to guide the eyes.
The sizes of the initial tensile velocity $v_{z0}$, 100, 200, 400 and 1000, are shown in the legend. The unit is m/s.
An enlarge portion of the curve for $v_{z0}=100$m/s is shown in the inset.
The growth of cavity can be described by the following stages: (i) initial slow growth stage, (ii) linear growth stage which ends when the global rarefaction wave arrives at the upper free surface, (iii) slower growth stage which ends when the reflected compression wave arrives at the cavity, (iv) quicker growth stage and (v) linear growth stage.
Figure 6(b) shows the evolutions of the cavity dimensions in Horizontal(H) and Vertical(V) directions. The points are simulation results and the lines are plotted to guide the eyes.
It is interesting to observe that the growth in horizontal direction is quicker than that in vertical direction. Such a mechanism is equivalent to the ``necking effect" in macroscale. There exists also a linear stage in the growths of cavity dimensions. The growth rates increase with increasing the strength of tensile loading.
Figure 6(c) shows the initial linear growth rate of cavity volume versus initial strength of tensile loading $v_{z0}$. The points are for the slopes of fitting lines in Fig. (a) for the first linear growth stage, and the line are linear fitting result for the points.
It is clear that within the checked range the volume growth rate linearly increases with the initial tensile velocity $v_{z0}$.

Figures.7(a) and 7(b)  show the density fields of the material at two times, 7.2ns and 12ns. Figures 7(c) and 7(d) shows the corresponding pressure fields. The lower boundary of the cavity gradually becomes planar and parallel to the rigid wall.

%%%%%%%%%%%%%%%%%%%%%%%%%%%%%%%%%%%%%%%%
\begin{figure}[tbp]
\center\includegraphics*%
[bbllx=0pt,bblly=0pt,bburx=554pt,bbury=609pt,angle=0,width=0.85\textwidth]
{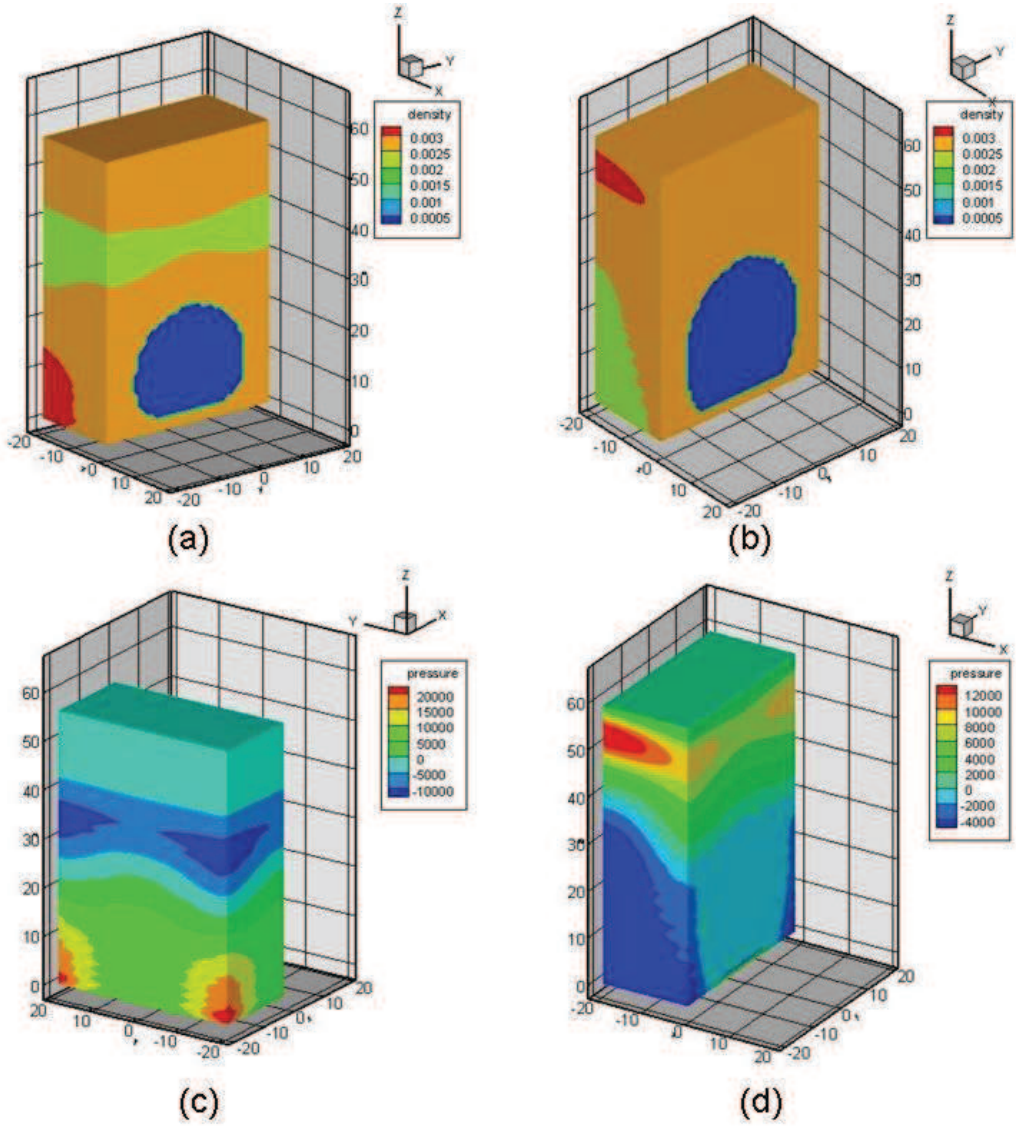}
\caption{Configurations with density field [(a) and (b)] and configurations with pressure field [(c) and (d)] at two times, 7.2ns and 12ns. Only the portion with $-20 \leq x \leq 0$ is shown in each plot.}
\end{figure}
%%%%%%%%%%%%%%%%%%%%%%%%%%%%%%%%%%%%%%%%%
%%%%%%%%%%%%%%%%%%%%%%%%%%%%%%%%%%%%%%%%
\begin{figure}[tbp]
\center\includegraphics*%
[bbllx=0pt,bblly=0pt,bburx=552pt,bbury=600pt,angle=0,width=0.85\textwidth]
{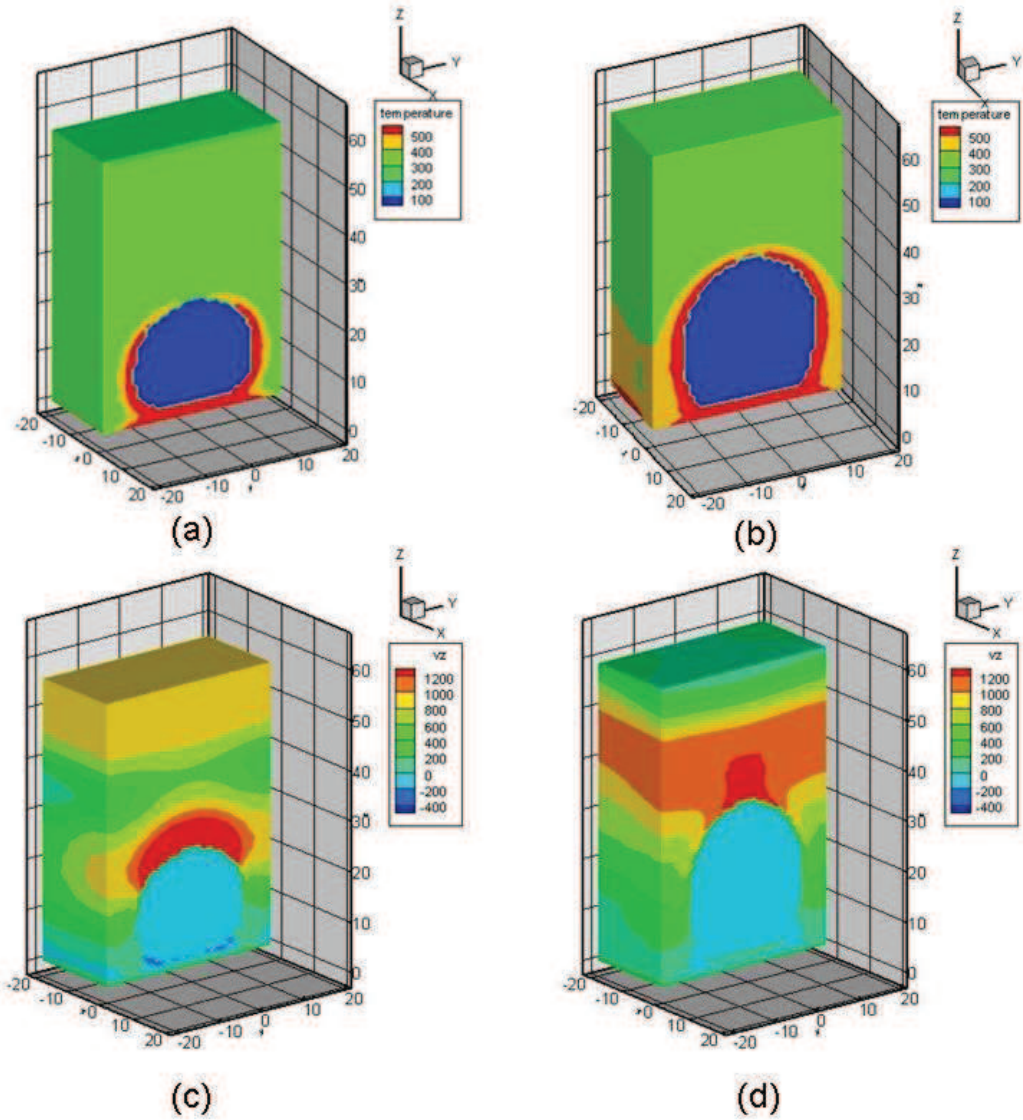}
\caption{Configurations with temperature field [(a) and (b)] and configurations with $v_z$ field [(c) and (d)] at two times, 7.2ns and 12ns. Only the portion with $-20 \leq x \leq 0$ is shown in each plot.}
\end{figure}
%%%%%%%%%%%%%%%%%%%%%%%%%%%%%%%%%%%%%%%%%

%%%%%%%%%%%%%%%%%%%%%%%%%%%%%%%%%%%%%%%%
\begin{figure}[tbp]
\center\includegraphics*%
[bbllx=0pt,bblly=0pt,bburx=579pt,bbury=604pt,angle=0,width=0.6\textwidth]
{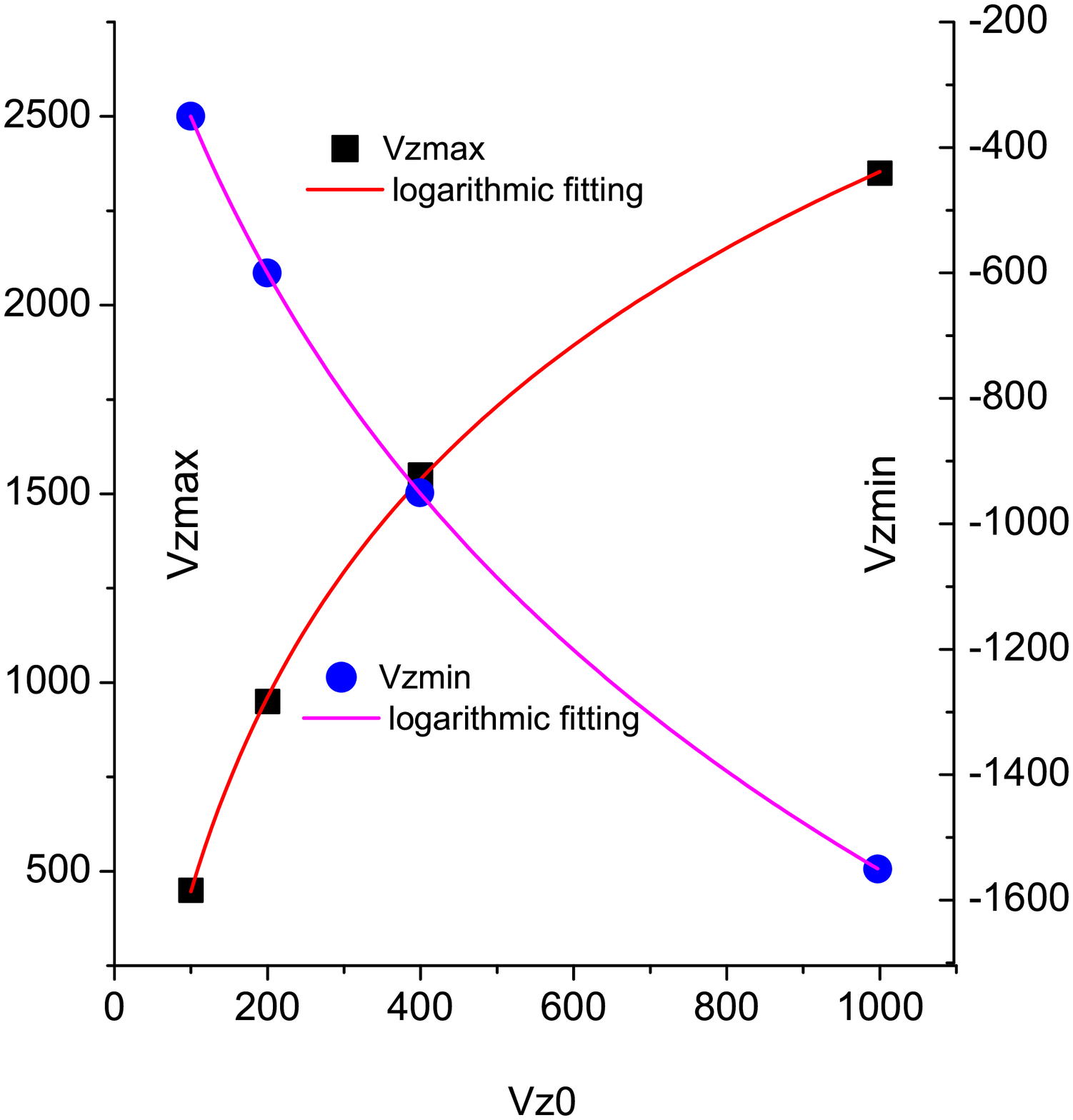}
\caption{Maximum upward particle velocity and maximum downward particle velocity versus initial tensile velocity $v_{z0}$. The points are simulation results and the lines are logarithmic fitting results.}
\end{figure}
%%%%%%%%%%%%%%%%%%%%%%%%%%%%%%%%%%%%%%%%%

\subsection{Energy transformation versus tensile strength}
For the case with uniform material, during the tensile loading, kinetic energy of the material transforms gradually to elastic potential energy and plastic work. Those energies distribute uniformly in planes parallel to the rigid wall. Although the material is three-dimensional, the dynamical and thermodynamical process is in fact one-dimensional. But for the case with cavities, the situation becomes much more complex. Figures 8(a) and 8(b) show the configurations with temperature field at the same two times as in Fig.7, from which, besides the cavity morphology, we can understand better the energy transformation from kinetic to thermal. There is a high temperature layer surrounding the deformed cavity. That is because the plastic work by the stresses is pronounced in that region. Figures 8(c) and 8(d) show the configurations with $v_z$ field at the same two times. With the reflecting back of compression wave from the upper free surface, the distribution range of high particle velocity becomes narrower.

Figure 9 shows the maximum upward particle velocity, $v_{z\max}$,  above the cavity and maximum downward particle velocity, $v_{z\min}$, below the cavity versus the initial tensile velocity $v_{z0}$. The points are simulation results and the lines are fitting results. It is interesting to observe that both $v_{z\max}$ and $|v_{z\min}|$ logarithmically increase with the initial tensile velocity $v_{z0}$.

\section{Conclusions}

A three-dimensional material point simulation study on cavity growth in metal materials subjected to dynamic loading is conducted. Interactions of rarefaction wave with an existing cavity and the ultimate interactions of the cavity with its periodic images are carefully investigated.
During the tensile loading procedure, some material particles below the cavity show high speeds in the opposite direction. Within the region subjected to the global rarefaction wave some local regions may show positive pressures. Neighboring cavities get interaction via the coalescence of isobaric contours. The deformation of cavity under tensile loading shows staged behaviors. After the initial slow growth stage, the volume and the dimensions in both the tensile and transverse directions show linear growth rate with time until the global tensile wave reaches the upper free surface. The growth rate in the tensile direction is slower than that in the transverse direction. The volume growth rate  linearly increases with the initial tensile velocity. After the global tensile wave passed the cavity, the maximum particle velocity in the tensile direction and the maximum particle velocity in the opposite direction increase logarithmically with the initial tensile speed. The shock wave reflected back from the cavity and compression wave from the free surface induce the initial behavior of interfacial instabilities such as the Richtmyer-Meshkov instability, which is mainly responsible for the irregularity of the cavity morphology. Temperature and distribution of hot spots are determined by the plastic work. Compared with the dynamical process, the heat conduction is much slower.

\section*{Acknowledgements}
We warmly thank Profs. Hongliang He, Ping Li and Xiaoyang Pei for helpful discussions. This work is supported by Science Foundation of State key Laboratory of Explosion Science and Technology[under Grant No. KFJJ14-1M], Science Foundation of China Academy of Engineering Physics
 [under Grant Nos. 2012B0101014 and 2011A0201002] and National Natural Science Foundation of China [under
Grant Nos. 11075021 and 11171038].

%%%%%%%%%%%%%%%%%%%%%%%%%%%%%%%%%%%%%%%%%%%%%%%%%%%%%%%%%%%%%%%
%\section*{References}


\begin{thebibliography}{99}

\bibitem{1}
G. T. Gray, P. J. Maudlin,  L. M. Hull,  Q. K. Zuo,  S. R. Chen,  Journal of Failure Analysis and Prevention, 2005, \textbf{5} (3): 7-17.

\bibitem{42} M. M. Carroll, A. C. Holt,  J. Appl. Phys. 1972, \textbf{27}(3): 1626-1636.

\bibitem{43} J. N. Johnson,  J. Appl. Phys. 1981 \textbf{52}: 2812 - 2825.

\bibitem{Becker1987} R. Becker, J. Mech. Phys. Solids 1987, \textbf{35}: 577.

\bibitem{44} M. Ortiz, A. Molinari,  J. Appl Mech. 1992, \textbf{59}: 48-53.

\bibitem{45} D. J. Benson, J. Mech. Phys. Solids. 1993, \textbf{41}(8):1285-1308.

\bibitem{46} X. Y. Wu, K. T. Ramesh, J. Mech. Phys. Solids 2003, \textbf{51}(1): 1-26.

\bibitem{Pardoen1998} T. Pardoen, I. Doghri, and F. Delannay, Acta Mater. 1998, \textbf{46}(2): 541 - 552.

\bibitem{Pardoen2000} T. Pardoen and J. W. Hutchinson, J. Mech. Phys. Solids 2000, \textbf{48}(12): 2467-2512

\bibitem{Orsini2001} V.C.Orsini and M.A.Zikry, Int. J. Plast. 2001, \textbf{17}(10): 1393-1417.

\bibitem{2002a} V. Tvergaard and J. W. Hutchinson, Int. J. Solids Struct. 2002, \textbf{39}(13-14): 3581-3597.
.
\bibitem{2002b} T. I. Zohdi, M. Kachanov, and I. Sevostianov, Int. J. Plast. 2002, \textbf{18}: 1649.
\bibitem{27} D. R. Curran, L. Seaman, D. A. Shockey, Phys. Rep. 1987, \textbf{147}:253.

\bibitem{47} E. T. Seppala, J. Belak,  Phys. Rev. Lett. 2004, \textbf{93}: 245503.

\bibitem{48}
A.K. Zurek, W.R. Thissell, J.N. Johnson, D.L. Tonks, R. Hixson,
Journal of Materials Processing Technology, 1996, \textbf{60}(1-4): 261-267.

\bibitem{Zurek1998}
A. K. Zurek, J. D. Embury, A. Kelly, W. R. Thissell, R. L. Gustavsen, J. E. Vorthman, and R. S. Hixson,
AIP Conf. Proc. 1998, \textbf{429}: 423-426; doi:http://dx.doi.org/10.1063/1.55658 (4 pages)

\bibitem{49}
D. L. Tonks, A. K. Zurek, and W. R. Thissell
AIP Conf. Proc. 2002, \textbf{620}: 611-614; doi:http://dx.doi.org/10.1063/1.1483613 (4 pages)

\bibitem{50}
J.P. Bandstra, D.M. Goto, D.A. Koss, Materials Science and Engineering:  A 1998, \textbf{249}:46-54.

\bibitem{51} J.P. Bandstra, D.A Koss,  Materials Science and Engineering: A 2001, \textbf{319-321}:490-495.

\bibitem{52} J.P. Bandstra, D.A. Koss, A. Geltmacher, P. Matic, R.K. Everett, Materials Science and Engineering: A 2004, \textbf{366}:269-281.

\bibitem{53}
M.F. Horstemeyer, M.M. Matalanis, A.M. Sieber, M.L. Botos,
International Journal of Plasticity 2000, \textbf{16}(7¨C8): 979¨C1015.

\bibitem{PRB2005} E. T. Seppl\"{a}l\"{a}, J. Belak and R. E. Rudd, Phys. Rev. B 2005, \textbf{71}: 064112.

\bibitem{MSMSE2006} L. M. Dupuy and R. E. Rudd, Modelling Simul. Mater. Sci. Eng. 2006, \textbf{14}: 229.

\bibitem{Pang1} W. Pang, G. Zhang, Aiguo Xu, G. Lu, Chin. J. Comp. Phys. 2011, \textbf{28}: 540-546. (in Chinese)

\bibitem{Pang2}  W. Pang, P. Zhang, G. Zhang, Aiguo Xu,, X. Zhao,  Science China: Phys. Mech. \& Astron. 2012, \textbf{42}: 464-474.

\bibitem{MPM1} D. Burgess, D. Sulsky, J. U. Brackbill, J. Comput. Phys. 1992, \textbf{103}(1): 1-15.

\bibitem{MPM2} S. Bardenhagen, J. Brackbill, and D. Sulsky, Comput. Methods Appl. Mech. Eng. 2000, \textbf{187}(3-4): 529-541.

\bibitem{MPM4} N. P. Daphalapurkar, H. Lu, D. Coker, R. Komanduri, Int. J. Fract. 2007, \textbf{143}(1): 79-102.

\bibitem{MPM5} S. Ma, X. Zhang, X. M. Qiu, Int. J. Impact Eng. 2009, \textbf{36}(2): 272-282.

\bibitem{CTP2008} X. F. Pan, Aiguo Xu, G. Zhang, et al, Commun. Theor. Phys. 2008, \textbf{49}(5): 1129-1138.

\bibitem{JPD2008} X. F. Pan, Aiguo Xu, G. Zhang and J. Zhu,  J. Phys. D: Appl. Phys. 2008, \textbf{41}: 015401.

\bibitem{JPCM2007} Aiguo Xu, X. F. Pan, G. Zhang  and J. Zhu, J. Phys.: Condens. Matter 2007, \textbf{19}: 326212.

\bibitem{CTP2009} Aiguo Xu, G. Zhang, X. F. Pan, and J. Zhu, Commun. Theor. Phys. 2009, \textbf{51}(4): 691-699.

\bibitem{CTP2009b} Aiguo Xu, G. Zhang, P. Zhang, X. F. Pan, and J. Zhu, Commun. Theor. Phys. 2009, \textbf{52}(5): 901-908.

\bibitem{SciCN2010} Aiguo Xu, G. Zhang, H. Li, Y. Ying, X. Yu, and J. Zhu, SCIENCE CHINA: Physics, Mechanics \& Astronomy 2010, \textbf{53}(8): 1466-1474.

\bibitem{PhysScr2010} Aiguo Xu, G. Zhang, Y. Ying, P. Zhang and J. Zhu,  Phys. Scr. 2010, \textbf{81}: 055805.

\bibitem{CAMWA2011} Aiguo Xu, G. Zhang, H. Li, Y. Ying, J. Zhu,  Computers and Mathematics with Applications 2011, \textbf{61}(12): 3618-3627.


\bibitem{CModel} F. Auricchio, L. B.  da Veiga, Int. J. Numer. Meth. Engng. 2003, \textbf{56}(10): 1375-1396.

\bibitem{explosion} B. Zhang, et al. \textit{Explosion physics}, Ordance Industry Press of China, Beijing, 1997.

\end{thebibliography}
\end{document}